\documentclass[intlimits,twoside,a4paper]{article}
\usepackage{amsmath,amssymb}
\usepackage{graphicx}
\usepackage{dcolumn}% Align table columns on decimal point
\usepackage{subfigure}

\usepackage{color}

\usepackage[T2A]{fontenc}
\usepackage[cp1251]{inputenc}

\usepackage[eqsecnum]{cmpj2}
\issue{2016}{19}{2}{23803}
\doinumber{10.5488/CMP.19.23803}

\title[An effective Hamiltonian approach for D-B-A  electronic transitions]
{An effective Hamiltonian approach for Donor-Bridge-Acceptor  electronic transitions: Exploring the role of bath memory}
\author[E.R. Bittner]{E.R. Bittner}
\address{Department of Chemistry, University of Houston, Houston TX 77204, USA}

\authorcopyright{E.R. Bittner, 2016}
\date{Received November 30, 2015, in final form January 14, 2016}

\begin{document}
\maketitle
\begin{abstract}
We present here a formally exact model for electronic transitions between an initial (donor) and final (acceptor) states linked
by an intermediate (bridge) state.  Our model incorporates a common set of vibrational modes that are coupled to
the donor, bridge, and acceptor states and serves as a dissipative bath that destroys quantum coherence between the
donor and acceptor.    Taking the memory time of the bath as a free parameter, we calculate transition rates for a heuristic 3-state/2 mode
Hamiltonian system parameterized to represent the energetics and couplings in a typical organic photovoltaic system.   Our results indicate that
if the memory time of the bath is of the order of 10--100~fs,
a two-state kinetic (i.e., incoherent hopping) model will grossly underestimate overall  transition rate.
 \\ charge transfer
\pacs
87.15.ht, % Ultrafast dynamics; charge transfer
82.20.Xr, % Quantum effects in rate constants (tunneling, resonances, etc.)
82.39.Jn % Charge (electron, proton) transfer in biological systems
\end{abstract}

\section{Introduction}

In  a multi-state system, quantum transitions from an initial to a final state are rarely direct
and often involve a coherent transfer between one or more intermediate states.  Moreover,
for transitions involving the electronic states of molecular systems in which there is significant coupling to a
large number of molecular vibrational degrees of freedom, one needs to properly
account for the the effects of dissipation, memory, and coherence \cite{tamura:107402,tamura:021103,pereverzev:104906,Bittner:2014aa}.
This is well known, and a number
of methods and approximations with varying degrees of exactness have been developed over the
years \cite{Stuchebrukhov:1994uq,medvedev:3821,scholes-arp-2003,doi:10.1146/annurev.bb.21.060192.002025}.
Two limits of approximations are the super-exchange model whereby population is transferred
from an initial (donor) state to an intermediate state or bridge before being transferred to the final (acceptor) state
and the hopping model where population is transferred in a sequence of discrete steps with no
quantum coherence between each subsequent step. Reference \cite{Wenger:2011} provides a succinct account of
recent progress in this area.

According to the super-exchange model, the electronic coupling between a donor and acceptor linked by
$n$ bridging units goes as
\begin{eqnarray}
H_\text{DA} = \frac{h_\text{DB}}{\Delta \epsilon} \left( \frac{h_\text{BB}}{\Delta \epsilon}\right)^{n-1} h_\text{BA},
\label{superex}
\end{eqnarray}
where $h_\text{DB}$ and $h_\text{BA}$ are the electronic (diabatic) couplings between $\text{D}\to \text{B}$ and from $\text{B}\to \text{A}$.
The energy gap $\Delta\epsilon$ is the tunneling gap taken as the difference between the $\text{D}\to \text{A}$ transition state
and the energy of the bridge-localized states, and $n$ is the number of bridging units with inter-bridge coupling $h_\text{BB}$.
Given $H_\text{DA}$, one can compute the transfer rates using the semi-classical Marcus equation
$$
k_\text{Marcus} = \frac{2\pi}{\hbar}|H_\text{DA}|^{2} \frac{1}{\sqrt{4 \pi k_\text{B}T \lambda}}\re^{-(\Delta G_{0} + \lambda)^{2}/4 \lambda k_\text{B}T},
$$
where $\Delta G_{0}$ is the driving force and $\lambda$ is the reorganization energy  \cite{Marcus:1985fk,marcus:679,marcus:966}.
Considerable amount of work has gone into establishing the validity of the super-exchange model
in models of hole-transport in DNA oligomers and electron-transfer dyads linked by $\pi$-conjugated bridges
\cite{Lewis:2005,Lewis:2005b,Starikov:2005,Lewis:2004,Wang:2004,Lewis:2003,Lewis:2581}.

In a separate context,  coherent long-range quantum transport may account for recent observations of
ultrafast exciton dissociation in organic polymer: fullerene based photovoltaic systems \cite{Provencher:2014aa,Gelinas:2013fk,Grancini:2013uq,Jailaubekov:2013fk,Troisi2013,ja1045742,tamura:107402,tamura:021103,bittner:034707,yang:045203}, in artificial light-harvesting systems \cite{Rozzi:2013fk}, and
natural photosynthetic systems \cite{1367-2630-12-6-065042}.
However, the difficulty using equation~(\ref{superex}) to compute matrix elements is that the intermediate dynamics within
the bridge are completely excluded and it does not account for the fact that the bridge itself may have
discrete vibronic states.

In the case of organic photovoltaics,  current debate concerns whether or not the charge separation
occurs via coherent tunneling or incoherent hopping mechanisms.  Pump-push-probe experiments \cite{Bakulin16032012,Gelinas:2013fk} and time-resolved resonant Raman experiments \cite{Provencher:2014aa}
tend to support the notion that that delocalized charge-transfer states are precursors for the formation of
charge-separated or polaron states, a view supported by  recent theoretical work \cite{Bittner:2014aa} and others \cite{Yao:2015ab,Rozzi:2013fk}.  On the other hand, entropic effects and randomness would lead to localized states
and an incoherent hopping mechanism.

With this in mind, we set about to construct a suitable super-exchange theory
that accounts for a common vibronic bath coupled to all the electronic states involved in the system and accounts for
the fact that the longer the population remains in the bridging state, quantum decoherence will effectively kill the
coherent transfer between $\text{D}\to \text{A}$.  We also desire a model that can take input directly from a
quantum chemical evaluation of the diabatic potentials  and electronic
couplings for realistic molecular systems \cite{subotnik2010predicting,xunmo1,doi:10.1021/jp503041y}.
Our approach is similar to that developed recently by Voityuk to study electron transport on
molecular wires \cite{C2CP40579B}.   As test case, we consider the interstate relaxation dynamics in a
model for charge-separation in an organic heterojunction system.

\section{\label{sec:level1}Theoretical approach}

We consider here a model  system consisting of three diabatic
electronic states denoted as $\text{D}$, $\text{B}$, and $\text{A}$ for ``donor'', ``bridge'' and ``acceptor'' corresponding to the electronic configurations:
\begin{eqnarray}
|\text{D}\rangle = | \text{D}^{*}\cdots \text{B} \cdots \text{A} \rangle, \\
|\text{B}\rangle = |\text{D}\cdots \text{B}^*\cdots \text{A}\rangle,  \\
|\text{A}\rangle = |\text{D}\cdots \text{B}\cdots \text{A}^* \rangle,
\end{eqnarray}
where $*$ denotes which of the three electronic states is occupied.  The scenario is
ubiquitous for charge and energy transfer dynamics and we shall develop our theory without reference to
a specific physical process.  It suffices to say, $\text{D}$, $\text{B}$, and $\text{A}$ are simply different diabatic or localized electronic
states of the given physical system.
With respect to a  common origin, these have energies
$E_\text{D}$, $E_\text{B}$ and $E_\text{A}$ respectively.
We define a {\em common } set of phonon/vibrational modes using boson operators $[a_q,a_{q'}^\dagger] = \delta_{qq'}$
and frequencies $\omega_q $.
We shall use these
to define diabatic potentials
\begin{eqnarray}
H_0 = \sum_n|n\rangle \langle n|\left\{
\left[E_n + \sum_q g_{nq}(a_q^\dagger + a_q)\right]
+ \sum_q \hbar\omega_q a_q^\dagger a_q\right\}\nonumber \\
\end{eqnarray}
and off-diagonal electronic couplings
\begin{eqnarray}
V = \sum_{nm}\gamma_{nm}|n\rangle\langle m |
\end{eqnarray}
that do not depend upon the phonon variables.
Note that our linear coupling between the
electronic and phonon variables produces a linear displacement in
the origins of each electronic potential.  Our total Hamiltonian is then the sum $H = H_0 + V$.

We next want to remove the linear coupling terms by
performing a polaron (shift) transformation using
$$
\tilde H  = \re^{-S}H \,\re^S,
$$
where $S$ is an anti-Hermitian operator chosen such that
$$
[H_0,S] = -V.
$$
For our purpose at hand, we take
\begin{eqnarray}
S = \sum_{nq} |n\rangle\langle n | \xi_{nq}(a_q^\dagger - a_q)
\end{eqnarray}
and define our transformed operators using the equations of motion
method
$$
a_q(\tau)  = \re^{-\tau S}a_q \re^{\tau S},
$$
where $a_q(0) = a_q$   and $a_q(1) = \tilde a_q$. This
yields
$$
a_q(\tau) = a(0) - \tau\sum_n \xi_{nq}|n\rangle\langle n|.
$$
Similarly for $\tilde a_q^\dagger$.
Introducing these into $\tilde H$, one can diagonalize
each term with respect to the phonon variables by
setting $\xi_{nq} = g_{nq}/\hbar\omega_q$
\begin{eqnarray}
\tilde H_0 = \sum_n |n\rangle\langle n|\left(
\tilde E_n + \sum_q \hbar\omega_q a_q^\dagger a_q \right),
\end{eqnarray}
where $\tilde E_n$ is the new energy origin of the $n$th state
$$
\tilde E_n = E_n - \sum_q \frac{g_{nq}^2}{\hbar\omega_q}.
$$
Similarly, the off-diagonal (diabatic) coupling can be transformed
as follows:
$$
\tilde V = \sum_{nm}\gamma_{nm}
|n\rangle
\re^{\sum_q(\xi_{mq}-\xi_{nq})(a_q^\dagger-a_q)} \langle m |.
$$
We write the Hamiltonian operator as a $3\times 3 $ matrix in the
basis of the electronic states:
\begin{eqnarray}
\tilde H = \left[
\begin{array}{ccc}
\tilde H_\text{D}      &     \tilde V_\text{DB}   &  0 \\
\tilde V_\text{BD}      &     H_\text{B}   & \tilde V_\text{BA} \\
0      &     \tilde V_\text{AB}   &  \tilde H_\text{A}
\end{array}
\right].
\end{eqnarray}

We now seek to eliminate the phonon variables
and the explicit description of the bridge electronic state.
To this end, we use the Feschbach technique to define the projection operators
$
\hat P = |\text{D}\rangle \langle \text{D}| + |\text{A} \rangle \langle \text{A}|$
and
$\hat Q = 1 - \hat P = |\text{B}\rangle\langle \text{B}|$ and use these to
derive the Green's function for propagation within the $P$ subspace as
influenced by the dynamics within $Q$.
First, we write $H_P = \hat P \tilde H \hat P$,
$H_Q = \hat Q \tilde H \hat Q$,
and $H_{PQ} = \hat P \tilde H \hat Q$.
We then define an effective Hamiltonian by
formally solving the Schr\"odinger equation
within the $Q$ subspace and introducing this back into the
Schr\"odinger equation for the $P$ subspace,
$$
H_{\textrm{eff}} = H_P +H_{PQ}(z - H_{Q})^{-1} H_{QP},
$$
where $z$ is the energy taken as a continuous and for now, complex
variable.  $H_\text{eff}$ can be contracted to a $2\times 2$ block
matrix with elements
\begin{eqnarray}
H_{\textrm{eff}} = \left[
\begin{array}{cc}
\tilde H_\text{D}  + \Gamma_\text{DD}(z)   & \Gamma_\text{DA}(z) \\
\Gamma_\text{DA}^{*}(z)                  & \tilde H_\text{A} + \Gamma_{AA}(z)
\end{array}
\right],
\end{eqnarray}
where $\Gamma_{ij}(z)$ are the renormalized
diabatic couplings:
\begin{eqnarray}
\Gamma_{ij}^{(\pm)}(z)  = \tilde V_{i\text{B}}\left(z-\tilde H_\text{B} \pm \ri\eta\right)^{-1}\tilde V_{\text{B}j},
\end{eqnarray}
where we have introduced a small $\pm \ri \eta$ to insure a proper causality.

\subsection{Renormalized couplings}

We next derive expressions for these renormalized couplings.
To do so, we consider the matrix elements
between two vibrational configurations
\begin{eqnarray}
\langle \{n\} | \Gamma_{ij}^{(\pm)}(z) |\{m\}\rangle
= \langle \{n\} |\tilde V_{i\text{B}}(z-\tilde H_\text{B}\pm \ri\eta)^{-1}\tilde V_{\text{B}j}
|\{m\}\rangle,
\end{eqnarray}
where the kets $|\{m\}\rangle $ denote a state in the Fock space
specified by the occupation numbers of each mode
$$
|\{m\}\rangle = \prod_q | m_q \rangle.
$$
%\begin{widetext}
Since the renormalized diabatic couplings
do not couple between different phonon modes, we can
write these as follows:
\begin{eqnarray}
\langle \{n\} | \Gamma_{ij}^{(\pm)}(z) |\{m\}\rangle
 = \sum_q \langle n_q | \Gamma_{ij}^{(\pm)}(z) |m_q\rangle
\end{eqnarray}
and evaluate each term mode by mode
\begin{eqnarray}
\langle n_q | \Gamma_{ij}^{(\pm)}(z) |m_q\rangle
&=& \langle n_q |
\tilde V_{i\text{B}}(z-\tilde H_\text{B} \pm \ri\eta)^{-1}\tilde V_{\text{B}j}
 |m_q\rangle \nonumber
\\
&=&
\sum_{l_q}
\langle n_q |
\tilde V_{i\text{B}}
|l_q\rangle\langle l_q|
(z-\tilde H_\text{B} \pm \ri\eta)^{-1}
|l_q\rangle\langle l_q|\tilde V_{\text{B}j}
 |m_q\rangle \nonumber \\
&=&
\sum_{l_q}
\gamma_{i\text{B}}\gamma_{\text{B}j}
\langle n_q | \exp\big\{(\xi_{iq}-\xi_{Bq})(a_q^\dagger - a_q)\big\}| l_q\rangle \nonumber\\
&& \times \frac{1}{z-(\tilde E_\text{B} + \hbar\omega_q l_q \pm \ri\eta)}
\langle l_q | \exp\big\{(\xi_{Bq}-\xi_{mq})(a_q^\dagger - a_q)\big\}| m_q\rangle.
\end{eqnarray}
Define $\Delta\xi_{q}^{ij} = \xi_{iq} - \xi_{jq}$ and write this in terms of the
nuclear overlap integrals
$$
S_{lmq}^{ij}(\Delta\xi^{ij}_{q}) =
\langle l_q | \exp\big\{\Delta\xi^{ij}_{q}(a_q^\dagger - a_q)\big\}| m_q\rangle.
$$
These are then used to construct the matrix elements of the couplings
\begin{eqnarray}
\langle n_q | \Gamma^{(\pm)}_{ij}(z) |m_q\rangle
&=&
\sum_{l_q}
\gamma_{i\text{B}}\gamma_{\text{B}j}
\frac{S_{nlq}^{i\text{B}}S_{lmq}^{\text{B}j}}{z-(\tilde E_\text{B} + \hbar\omega_q l_q)\pm \ri\eta}.
\end{eqnarray}
This last expression can be understood in the following way.
An electron from the donor (in vibrational configuration ${\bf m} = \{{\bf m}_{1},{\bf m}_{2},\cdots\}$)
is first transferred to the bridging state and the system evolves within the manifold of
vibrational states as specified by the nuclear overlap factors between the
donor and the bridge.  This state then scatters into a specified
vibrational  of the acceptor state ${\bf n} = \{{\bf n}_{1},{\bf n}_{2},\cdots\}$
with probability dictated by the nuclear overlap between the
the bridge and the acceptor.
Using the properties of harmonic oscillator states, the overlaps can be evaluated exactly as follows:
\begin{eqnarray}
S_{nl}(\xi)&=&\frac{1}{\sqrt{n!l!}}\re^{-\xi^{2}/2}\sum_{s=0}^{n}
\left(
\begin{array}{c}
n \\ s
\end{array}
\right)
(-\xi)^{n-s}
\frac{\rd^{s}}{\rd\xi^{s}}(\xi^{s}).
\end{eqnarray}
Note that $\xi$ is related to the relative displacements between the harmonic wells
via $\xi =\sqrt{m\omega/\hbar} Q$. In this last expression, we have dropped the
explicit reference to the electronic states.

\subsection{Perturbation series}

To compute the transition probability from the donor to the acceptor states,
we construct the Green's function using the Dyson series:
\begin{eqnarray}
G_\text{DA}^{(\pm)}(z) =
 G_\text{D}^{(\pm)}(z)\Gamma_\text{DA}^{(\pm)}(z)G_\text{A}^{(\pm)}(z) + \cdots
\end{eqnarray}
which can be written as a sum over all combinations of interaction vertices represented as tadpole diagrams
where the loop denotes integration over the vibrational states of the bridge.
\begin{eqnarray}
\includegraphics[]{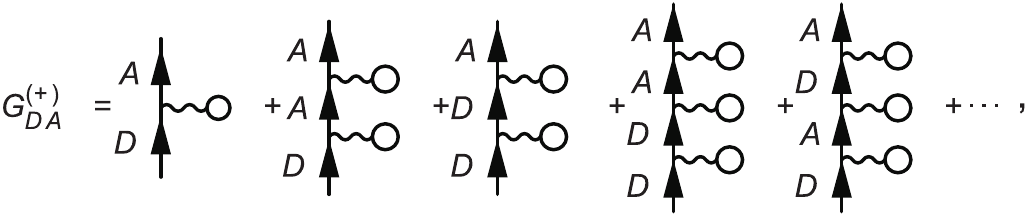}
\end{eqnarray}
where each plain propagator (vertical arrow) is the Green's function for either the donor or acceptor subspaces ($G_\text{D}^{(+)}$ or
$G_\text{A}^{(+)}$)  and the curly
line indicates an interaction with the bridge (loop) (i.e., $\Gamma_\text{DA}^{(+)}$).
To simplify the matters,  let us consider the plain propagator as being ``dressed'' by the interaction with the bridge and include only
vertices that scatter from $\text{D} \to \text{A}$ or $\text{A} \to \text{D}$.
In essence, we write
$$
{G}_\text{D,A}^{(\pm)} (z) = \frac{1}{z -\big[\tilde H_\text{D,A} + \Gamma_\text{D,A}(z)\big] \pm \ri\eta}
$$
as the dressed propagator within the $\text{D}$ or $\text{A}$ sub-spaces where $\Gamma_\text{D,A}(z)$ is
the self-energy of the donor or acceptor due to its interaction with the $\text{B}$ sub-space.
For the off-diagonal  terms, the terminal vertices must be ``D'' for the incoming and ``A'' for the outgoing terms,
implying we have an odd number of interaction vertices.
In what follows, we shall neglect the ``diagonal'' self-energy terms
and consider only the first term in the perturbation series.

\subsection{Golden rule transition rates}

We now use renormalized couplings to compute the transition rates from the donor to the acceptor states.
To this end, we recognize that the $\Gamma_{ij}(z)$ vertices are the Fourier-Laplace transforms of Heisenberg operators $\hat\Gamma_{ij}(t)$.
With this in mind, we can construct time-correlation functions using the Wiener-Khinchin theorem
$$
{ C}_\text{DA}(t) = \langle G_\text{DA}^{(+)}\star (G_{AD}^{(-)}) \rangle,
$$
where $\star$ denotes the convolution integral
\begin{eqnarray}
{ C}_\text{DA}(t) &=& \left\langle \int_{-\infty}^\infty \rd t\,\; G_\text{DA}^{(+)}(z)G_\text{AD}^{(-)}(z)\re^{\ri z t/\hbar}  \right \rangle
%\nonumber \\ &=&
= \int_{-\infty}^\infty \rd t \;\re^{\ri z t/\hbar} C_\text{DA}(z)
\end{eqnarray}
and $\langle \cdots \rangle$ denotes a sum over both initial and final states, weighted by the populations of the
initial states.
\begin{eqnarray}
C_\text{DA}(z) =
 \frac{2\pi}{\hbar}
\sum_q\sum_{n_q,m_q}\left(\frac{\re^{-\beta\hbar\omega_q n_q}}{1-\re^{-\beta\hbar\omega_q}} \right)
|\langle n_q | G_\text{DA}(z) |m_q\rangle|^2. \nonumber \\
\label{spect-dens}
\end{eqnarray}
Note that the thermal average is a sum over {\em all} vibrational
modes of the system since we have not distinguished between vibrational
modes localized on the donor, bridge, or acceptor units.
 The golden-rule transition rate can then  be determined by evaluating $C_\text{DA}(z)$ at the
transition energy between the initial and final states,  $z = \hbar\omega_\text{DA}$.

\section{Model calculations}

Having derived a relatively compact expression for the golden rule rate between the initial and final electronic states, we
can apply this approach to a variety of electronic transitions which are mediated by an intermediate or ``bridge'' state.
In essence, this is a very close to exact expression for the rate of super-exchange between the states in which the electronic transitions
are coupled to a manifold of vibrational states.
It is important to point out that all the electronic states in our model system are coupled to a {\em common}  vibronic bath.
Hence, our model includes a memory effect that may linger within the vibronic bath over the course of the transition.
An important simplification to the model is to partition the vibronic couplings such that the donor, bridge, and acceptor units
are coupled to vibrational modes localized on just those units.   The model at hand easily allows for this simplification.

As a test case, we consider a 3-electronic state system denoted as $|\text{D}\rangle$, $|\text{B}\rangle$, and $|\text{A}\rangle$ coupled to two
independent oscillator modes.  We choose our couplings, energetics, and Huang-Rhys parameters
according to typical ranges of these values in
conjugated organic chromophores and loosely consider the higher frequency vibrational mode to correspond to the in-plane $\text{C}=\text{C}$
stretching modes ($\hbar\omega_{1} = 0.15$~{eV}) and the lower-frequency mode to correspond to the out of plane torsional modes of such molecules
($\hbar\omega_{2} = 0.015$~{eV)    \cite{karabunarliev:10219,Karabunarliev03b,karabunarliev:5863,Karabunarliev00}.
Table~\ref{table1} in the appendix gives a summary of the
parameters used in our model, and  figure~\ref{figure1} shows the adiabatic electronic potentials along the line connecting the
local minima of the $\text{A}$ and $\text{B}$ diabatic potentials.   Briefly, we consider the transfer of population from an initially prepared ``donor'' state with a thermal population of the two vibrational modes.

\begin{figure}[!h]
\centerline{
\includegraphics[width=0.45\textwidth]{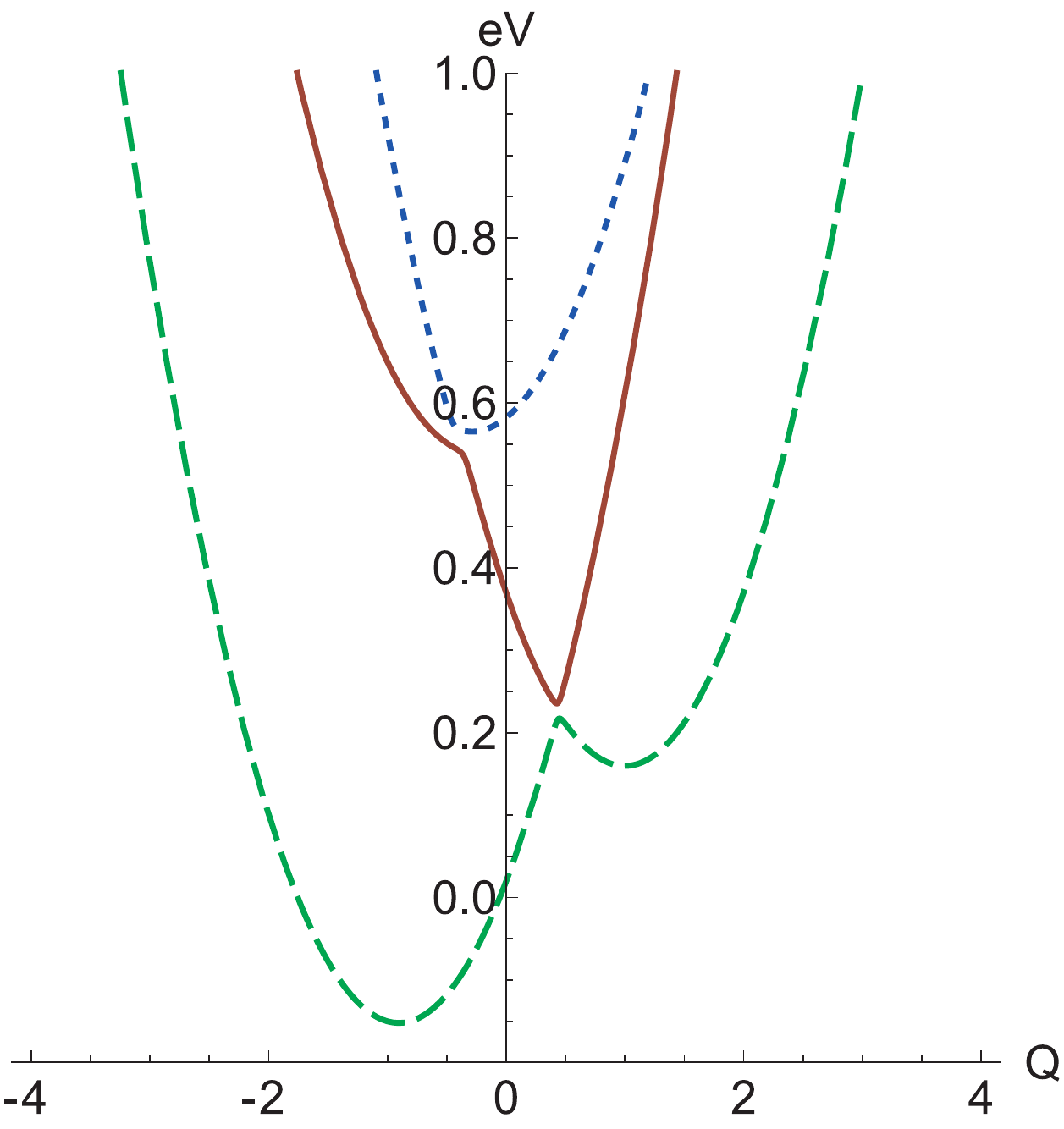}
}
\caption{(Color online) Model adiabatic potential curves along the coordinate connecting the bridge and the acceptor minima.}
\label{figure1}
\end{figure}

Formally, we introduced $\eta$ in our Greens function to preserve the causality when performing integrations over $z$.
 $G(z)$ is also a measure of the density of states since each pole represents the energy of a stationary state.
 For an isolated system, the poles of $G(z)$ become delta-functions in the density of states.  However,
 for a system in contact with a dissipative medium, each delta-function becomes a Lorenzian with an energy width, $\delta E$ and
lifetime $\tau$  as governed by the so-called ``time-energy'' uncertainty relation $\delta E \cdot  \tau \geqslant \hbar/2$.
 Thus, by introducing $\eta$ and shifting the poles off the real-energy axis, we introduce a natural life-time of $\tau \hbar/2\eta$ to each
 vibrational mode in our model.   Consequently, $\eta$ is a free parameter that can be used to characterize the
memory effects of the vibronic bath. In essence, making $\eta$ smaller, effectively {\em increases} the
amount of time the vibronic bath retains the memory of the initial state.

\begin{figure}[!t]
\begin{center}
\subfigure[]{
\includegraphics[width=0.48\columnwidth]{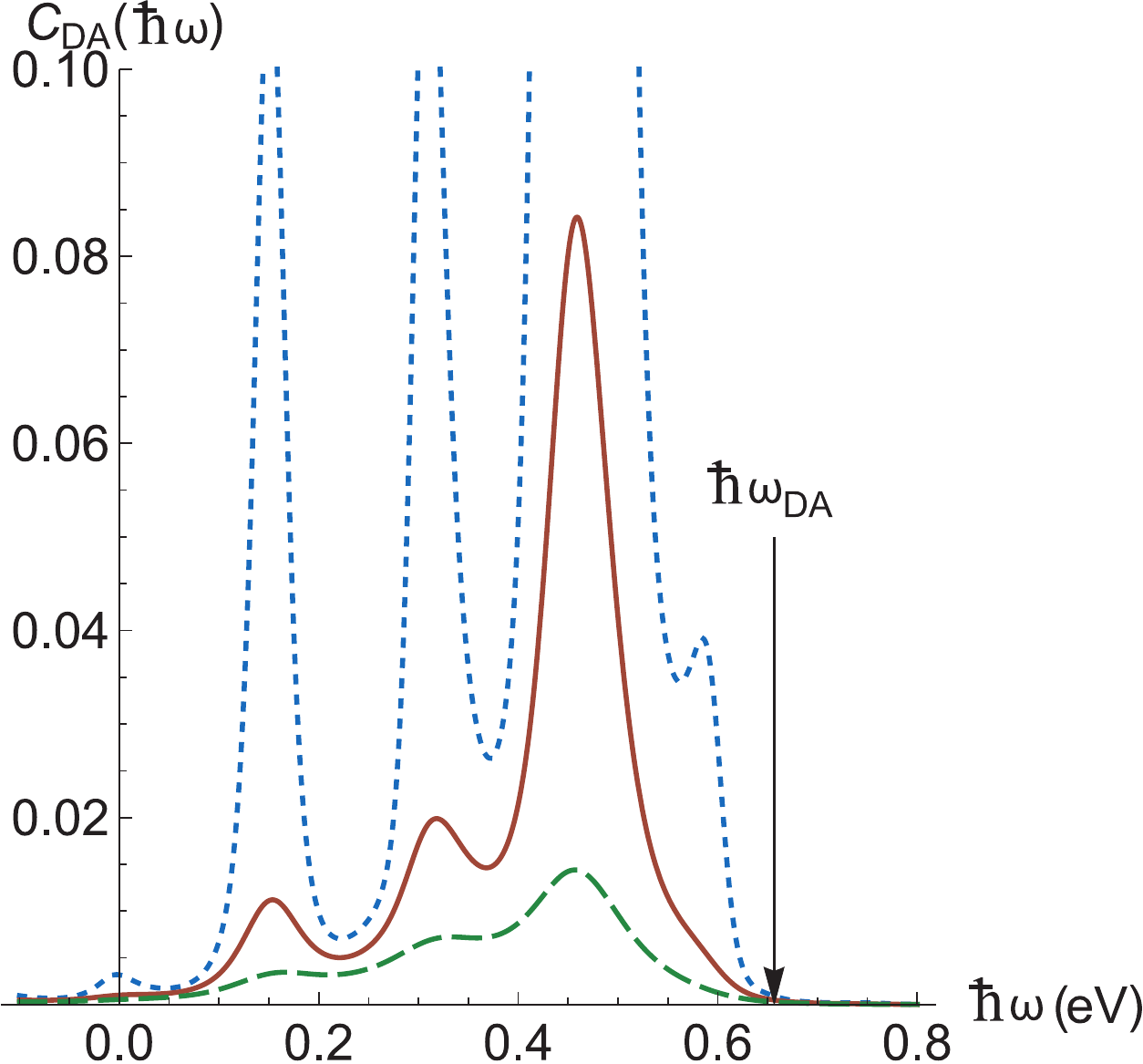}
}
\subfigure[]{
\includegraphics[width=0.48\columnwidth]{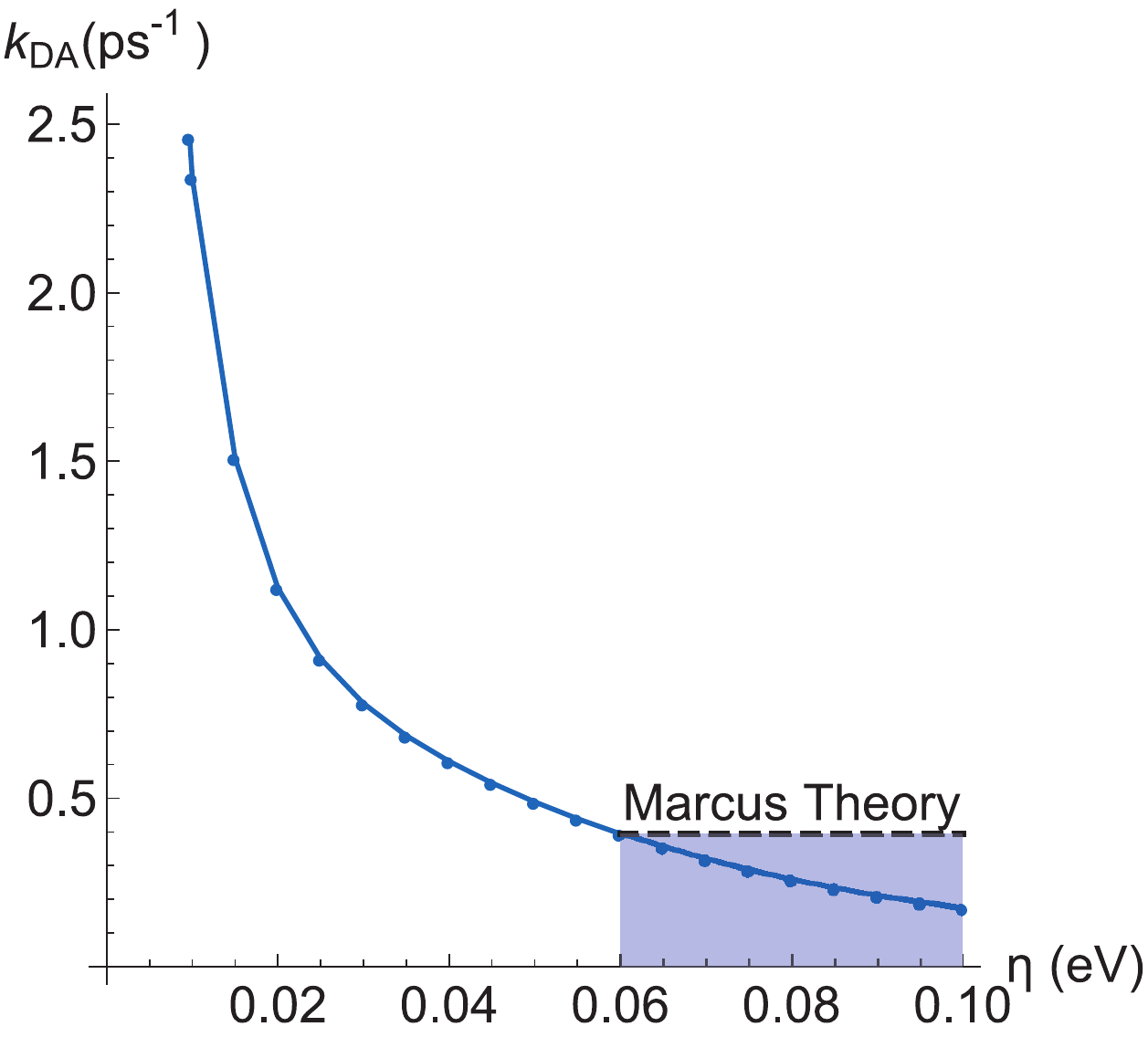}
}
\end{center}
\caption{(Color online)  (a) Spectral density for $\eta= 0.025$~{eV} (blue), $\eta= 0.05$~{eV} (gold), and $\eta= 0.075$~{eV} (green).
The arrow marks the position of the $\text{D}\to \text{A}$ transition frequency between the electronic minima of the $\text{D}$ and $\text{A}$ adiabatic potential wells.
(b) Computed golden-rule rate constants for model system as function of $\eta$.}
\label{figure2}
\end{figure}

Figure~\ref{figure2}~(a) shows the computed spectral density [equation~(\ref{spect-dens})] over a range of values of $\eta$.
 The peaks correspond to specific energies in which the coupling between the donor and acceptor is strongest, i.e.,
to resonances within the vibronic bath.   Decreasing $\eta$ causes the peaks to become increasingly narrow and more
peaked at the resonant frequencies.  However, since the golden-rule rate is evaluated at a frequency $\hbar\omega_\text{DA}$ that
may not be exactly on-resonance with the bath frequency, the changes in the spectral line-shape of the bath have a dramatic
impact on the $\text{D}\to \text{A}$ transition rate.   In figure~\ref{figure2}~(b), we give the computed $\text{D}\to \text{A}$ rates
as a function of $\eta$.    As expected, the rates decrease as we decrease the memory time of the bath.

As a point of comparison, we consider the extreme case where the $\text{D}\to \text{A}$ transition can be considered as a 3 step process of
$\text{D}\to \text{B} \to \text{A}$ with a {\em full} loss of coherence (memory) at each step with rate constants given by  Marcus theory.
Integrating the resulting kinetic equations gives an exact expression
$$
p_{A}(t) = 1 - \frac{k_{2} \re^{-k_{1}t} + k_{1}\re^{-k_{2}t}}{k_{2}-k_{1}}.
$$
 Taking the half-life of 2~ps to be characteristic of the $\text{D}\to \text{A}$ transition rate,
we can make a direct comparison with the DBA rate computed above and estimate
a lower limit of $\eta$ for which a two-step (non-superexchange) model would be appropriate.
For the case at hand, for $\eta < 0.06$~eV, the direct DBA (super-exchange) model predicts the rates faster
than the two-step (Marcus) model implying that for cases in which the memory time of the bath is longer than
a few 10's of fs,  sequential models of the kinetics will grossly underestimate the transition rates.

We can also approximate the DA coupling using the super-exchange model by taking the energy gap
to be the energy difference between where the $\text{D}$ and $\text{A}$ diabatic curves intersect and the energy minimum of $\text{B}$
as $\Delta \epsilon = - 0.75$~eV, and one obtains a very small super-exchange
coupling of $1.32\times 10^{-4}$~eV and a
vanishingly small super-exchange rate of $3.6 \times 10^{-7}~{\text{fs}}^{-1}$.   In this case, the neglect of the
vibronic coupling grossly underestimates the rate.

\section{Discussion}

We have presented here a model for computing the super-exchange transition rates between a model $\text{D} \to \text{B} \to \text{A}$ system
composed of 3 electronic states coupled to a common set of vibrational modes.   The analysis is generalizable to any number of
cases in which there is little to no direct electronic coupling  between the initial ($\text{D}$) and the final (A) diabatic states of the
system.     As a proof of concept, we analyze a simple 3-state/2-mode model parameterized as to represent a prototypical
organic photoexcitation system where the relaxation from the initial to the final state proceeds via some common intermediate state.
Comparing to a two-step model in which the intermediate rates are determined via Marcus  theory, indicates that one needs to be careful in considering  the memory effects in the vibrational bath when applying the Marcus theory
 to a sequence of electronic transitions that
occur on the ultra-fast timescale.
Our group is currently exploring using the approach delineated in this paper to examine the prompt generation of photocurrent
observed in polymer-fullerene based organic photovoltaic cells
\cite{Provencher:2014aa,Bittner:2014aa}.

\section*{Acknowledgements}

The author offers his hearty congratulations to
Prof. Haymet on the occasion of his first 60 successful trips around the sun.
The work at the University of Houston was funded in part by the
National Science Foundation (CHE-1362006)
and the Robert A. Welch Foundation (E-1337).

\appendix
\section{Model parameters}
\begin{table}[h]
\caption{Model parameters for 3-level/2 mode problem.\label{table1}}
\begin{center}
\begin{tabular}{|c|c|c|}
\hline\hline
Description       &  Symbol             & Value \\
\hline\hline
donor energy   & $E_\text{D}$    &  0.5 eV \\
bridge energy  & $E_\text{B}$     &  0.55 eV \\
acceptor energy  & $E_\text{A}$     & 0 eV \\
D-A Diabatic coupling & $\gamma_\text{DA}$  & $-0.01$~eV \\
A-B Diabatic coupling & $\gamma_\text{BA} $ & $-0.01$~eV \\
\hline
Electron Phonon Couplings   &$g_{ni}$  &  \\
donor state  & $(g_\text{D1},g_\text{D2})  $    & (0, 0) \\
bridge state  & $(g_\text{B1},g_\text{B2})  $   & (0.15, 0.06)~eV \\
acceptor state  & $(g_\text{D1},g_\text{D3})   $   & ($-0.15$, $0.01$)~eV \\
\hline
Adiabatic minima    &  $\tilde E_\text{D}$    & 0.5 eV\\
 			  &  $\tilde E_\text{B}$    &  0.16 eV\\
			   &  $\tilde E_\text{A}$    & $-0.1567$ eV\\

\hline
\hline
\end{tabular}
\end{center}
\end{table}%

%\bibliographystyle{unsrt}
%\bibliography{/Users/ebittner/Dropbox/References}

\ukrainianpart

\title{Підхід ефективного гамільтоніану для електронних переходів донор-місток-акцептор: дослідження \\ ролі пам'яті дисипативного середовища}
\author{Е.Р. Біттнер}
\address{Хімічний факультет, Університет Г'юстона, Г'юстон, Техас 77204, США}

\makeukrtitle

\begin{abstract}
\tolerance=3000%
Представлено формально точну модель електронних переходів між початковим (донор) та кінцевим (акцептор) станами, які зв'язані проміжним (місток) станом. Наша модель включає спільний набір коливних мод, які взаємодіють з донорним, містковим та акцепторним станами, та служить
як дисипативний термостат, що порушує квантову когерентність між донором і акцептором.
Беручи час пам'яті термостата як вільний параметр, ми розраховуємо інтенсивність переходів для евристичного 3-стани/2 модового гамільтоніана системи, параметризованого для опису енергетики та взаємодій в типово органічній фотовольтаїчній системі. Наші результати вказують, що якщо час пам'яті термостату є порядку 10--100~пс, дво-станова кінетична (тобто з некогерентним перескоком) модель значно недооціюнює загальну інтенсивність переходів.

\keywords супер-обмін, теорія електронного переходу, фотовольтаїка органічних сполук, надшвидка  динаміка, перескок заряду

\end{abstract}

\end{document}